# Implementing agile healthcare frame works in the context of low income countries: Proposed Framework and Review

**PK Dutta**

**School of Engineering and Technology, Amity University Kolkata, India**

pkdutta@kol.amity.edu

**Abstract**

Agile healthcare frameworks, adapted from the iterative and responsive methodologies of IT and manufacturing industries, present transformative opportunities for improving healthcare systems in low-income regions. This study examines the integration of Agile principles in resource-constrained environments, with Ghana serving as a focal case. The research identifies critical benefits, including enhanced adaptability, iterative planning, and stakeholder collaboration, which address systemic challenges such as infrastructure limitations, workforce shortages, and the "know-do gap." By leveraging digital tools like mobile health (mHealth) applications and the District Health Information Management System (DHIMS), Ghana's healthcare system demonstrates the scalability and efficacy of Agile methodologies in improving patient outcomes and resource allocation. Policy alignment, such as through Ghana's National Health Insurance Scheme (NHIS), emerges as essential for sustaining Agile practices. Key findings underscore Agile's ability to foster real-time decision-making, promote community engagement, and facilitate interdisciplinary collaboration. This paper highlights actionable strategies and systemic innovations that enable Agile frameworks to deliver equitable and high-quality care in resource-constrained environments. The implications extend beyond Ghana, offering a scalable model for other low-income regions striving to enhance healthcare delivery and resilience.

**Introduction**

In an era where healthcare demands are becoming increasingly complex and unpredictable, agility has emerged as a critical enabler for system adaptability and resilience. Agile methodologies, initially designed for the fast-paced and innovation-driven world of IT and manufacturing, have demonstrated transformative potential across industries. These methodologies prioritize iterative processes, responsiveness to change, and stakeholder collaboration, making them well-suited for addressing the dynamic challenges faced by healthcare systems. The evolving complexities of global healthcare demand adaptive, resilient systems capable of addressing both immediate challenges and long-term needs. Agile healthcare frameworks have emerged as transformative models, drawing inspiration from Agile methodologies in industries like IT and manufacturing. These frameworks emphasize iterative processes, stakeholder collaboration, and rapid response to changing conditions. By prioritizing flexibility and continuous improvement, Agile healthcare frameworks have the potential to revolutionize care delivery, particularly in environments characterized by resource constraints. For low-income and developing countries, where healthcare systems often grapple with inadequate infrastructure, workforce shortages, and financial limitations, the adoption of Agile principles offers a pathway to sustainable, high-quality care.

Critical to the success of Agile healthcare frameworks in poor countries are principles that directly address their unique challenges. Iterative planning and feedback loops ensure that care delivery adapts quickly to emerging needs, while transparency and collaboration foster trust among stakeholders, including patients, providers, and policymakers. Furthermore, Agile's focus on prioritizing value enables scarce resources to be allocated where they are most impactful. These principles align well with the requirements of healthcare systems in low-income regions, where demand for services often exceeds supply, and traditional hierarchical models struggle to cope with dynamic healthcare challenges. Agile methodologies tailored for low-income healthcare systems bridge the gap between evidence-based knowledge and its practical implementation, often referred to as the "know-do gap." By organizing workflows into manageable iterations or "sprints," healthcare teams can address critical needs promptly, evaluate outcomes, and adapt strategies based on real-time feedback. For example, deploying Agile practices in vaccination campaigns or maternal health initiatives has shown success in improving both efficiency and patient outcomes. Moreover, these methodologies encourage innovation by creating a culture of continuous reassessment and improvement, enabling even resource-limited systems to enhance care delivery while navigating economic and infrastructural constraints. This paper explores the integration of Agile healthcare frameworks in low-resource settings, focusing on their guiding principles, implementation strategies, and the transformative impact they can achieve in improving patient care and system efficiency. However, despite their front-line role in managing unpredictability, healthcare systems—especially in low- and middle-income countries (LMICs)—have been slow to adopt Agile principles. The healthcare landscape in resource-constrained settings is marked by chronic challenges, including limited infrastructure, workforce shortages, and financial constraints, which have been exacerbated by global crises like the COVID-19 pandemic. Nearly one-third of populations in LMICs reside more than two hours away from essential health services, and healthcare worker-to-patient ratios remain critically below the World Health Organization's recommended thresholds. These systemic deficiencies underscore the urgent need for frameworks that can drive efficiency, innovation, and resilience in these environments. Agile healthcare frameworks offer a promising solution by shifting from reactive to proactive strategies. By leveraging iterative cycles and customer-centric approaches, Agile principles can enable healthcare systems to anticipate potential issues, implement preventative measures, and continuously optimize care delivery. This transformation is not merely operational but also strategic, addressing the 'know-do gap'—the disparity between evidence-based knowledge and its application in practice. Bridging this gap is essential to improving patient outcomes and achieving equitable healthcare access.

This paper explores the potential of Agile frameworks to revolutionize healthcare delivery in poor countries. It highlights the critical Agile principles applicable to healthcare, examines methodologies for implementation in resource-limited settings, and discusses the challenges and benefits of adopting these frameworks. Through case studies and actionable strategies, we demonstrate how Agile methodologies can help healthcare systems overcome constraints and deliver sustainable, high-quality care. By addressing the unique needs of LMICs, this study seeks to contribute to the growing discourse on the application of Agile principles in non-traditional sectors, offering a roadmap for healthcare systems to adapt and thrive in the face of adversity.

In today's dynamic world, agility—a combination of balance, speed, strength, and coordination—has emerged as a critical factor in healthcare. It's no longer just about responding to change, but anticipating and acting ahead of it. Industries like IT and manufacturing have long embraced agility, rapidly adapting to market needs and innovating swiftly. In contrast, healthcare, despite being on the front lines of unpredictability, has lagged[7].

Healthcare can draw valuable insights from these agile industries. For example, the tech world's agile frameworks, with their iterative cycles and customer-centric approach, could be adapted to healthcare's unique environment. A shift from a reactive to a proactive mindset—anticipating potential health issues, implementing preventative measures, and continuously iterating care plans—will be a significant change[7].

## 1.1 Agile Methodologies in Low-Income Healthcare Systems

In the context of low-income healthcare systems, agile methodologies can bridge the significant gap between evidence-based knowledge and practice, often referred to as the 'know-do gap' [8][15]. Initially conceptualized by Rogers in 1962, the diffusion of innovation theory suggests that the perception of innovation, the adopters, and con- textual factors influence the rate of adoption [8][15]. These elements, often resistant to rapid change, can be addressed through agile frameworks that organize work in an iterative, predictable, and engaging manner, thereby promoting sustainable rapid innovation [8][15].

The dynamic nature of today's world demands agility—a blend of balance, speed, strength, and coordination. Industries such as IT and manufacturing have long harnessed the power of agility to adapt swiftly and innovate in response to market needs. Conversely, healthcare, despite its front-line role in managing unpredictability, has not kept pace [14]. By drawing on agile frameworks from these industries, which prioritize iterative cycles and a customer-centric approach, healthcare can shift from a reactive to a proactive mindset. This transformation involves anticipating potential health issues, implementing preventative measures, and continuously iterating care plans [14].

Healthcare policymakers in low-resource settings face unique challenges when implementing agile methodologies. Effective prioritization and resource allocation are crucial to ensure that the most critical areas are addressed first [14][15]. Agile and lean principles, while traditionally associated with software development, have proven effective in healthcare as well. These principles can help providers navigate the changing dynamics and economic pressures while meeting the rising expec- tations of patients and leveraging advancing technologies [16]. By adopting agile methodologies, healthcare systems in low-income countries can enhance efficiency and effectiveness, ultimately improving care delivery and patient outcomes [16].

1.2 Challenges of Implementing Agile in Poor Coun- tries

Implementing agile healthcare frameworks in low- and middle-income countries (LMICs) presents a unique set of challenges. One of the major obstacles is the existing gap in healthcare infrastructure and accessibility, which was further exac- erbated by the COVID-19 pandemic[13]. Before the pandemic, almost a third of the population in LMICs lived more than two hours away from essential healthcare services, and the ratio of healthcare workers to population was significantly below the World Health Organization's minimum recommendations[13]. The pandemic has worsened these issues, with the loss of healthcare workers and disruptions in transport systems due to lockdowns[13].

Effective governance and cross-sectoral partnerships are crucial to building resilient health systems that can adapt to emergencies and incorporate agile methodolo- gies[12]. For example, the experience from the COVID-19 pandemic has shown that countries need to take transformative actions to improve their health sector gover- nance and develop partnerships based on a One Health approach[12]. This approach emphasizes the importance of health service delivery and pandemic preparedness, prevention, and response[12].

Priority setting and resource allocation (PSRA) practices in high-income countries offer insights that could inform improvements in LMICs[10][11]. In high-income coun- tries, the process of making investment decisions is typically systematic, transparent, and evidence-informed[10]. Despite this, there is still room for improvement, and ongoing research aims to identify effective PSRA practices[11]. Applying similar systematic and transparent approaches in LMICs could enhance decision-making processes, ensuring that resources are prioritized and allocated effectively[10][11].

Margaret Kruk's work as chair of The Lancet Global Health Commission on High Quality Health Systems in the SDG Era underscores the importance of reviewing current knowledge, conducting empirical studies, and offering policy recommenda- tions to improve healthcare quality in LMICs[9]. Her efforts highlight the necessity for targeted strategies that address the unique challenges faced by these countries[9]. By focusing on these areas, policymakers in resource-poor settings can prioritize and allocate resources more effectively, ensuring that the most critical healthcare needs are addressed first.

## 2. Benefits of Agile Healthcare Frameworks in Poor Countries

Agile healthcare frameworks offer significant benefits in poor countries, where re- sources are often limited, and the demand for efficient and high-quality medical care is critical. Initially developed to expedite software development processes, Agile methodologies focus on quick value delivery through short sprints and iterative cycles, which can be adapted to healthcare to enhance patient outcomes and satisfaction[1].

In the dynamic world of healthcare, agility is key not just for responding to changes but for anticipating and acting proactively. This approach draws from successful practices in the IT and manufacturing sectors, which have long leveraged agile principles to adapt rapidly and innovate[2]. By adopting a proactive mindset, healthcare providers in impoverished regions can better anticipate potential health issues, implement preventative measures, and continuously iterate on care plans to improve health outcomes[2].

The COVID-19 pandemic has underscored the need for such adaptable frameworks. The crisis forced medical institutions worldwide to quickly adapt their workflows and processes, demonstrating the inherent flexibility and adaptability of Agile methodologies. These frameworks have helped healthcare professionals streamline operations, thus improving patient care and overall satisfaction, even under extreme pressure[3].

Agile's iterative and cooperative approach to project management emphasizes adaptability, client feedback, and continuous improvement. This is particularly vital in poor countries, where healthcare systems face the dual challenge of delivering high-quality care efficiently and innovating under constraints[4]. Despite its potential, only a minority of healthcare executives are currently familiar with Agile, though those who have implemented it report significantly better team performance compared to traditional methods[5]. The adoption of Agile healthcare frameworks in poor countries, such as Ghana, highlights the transformative potential of these methodologies in addressing systemic challenges and improving care delivery. Agile frameworks emphasize adaptability, collaboration, and iterative problem-solving, making them well-suited to the dynamic and resource-constrained environments typical of low-income nations. Ghana's healthcare system provides a compelling case study of how Agile principles can bridge gaps in access, infrastructure, and quality of care.

*2.1 Enhancing Flexibility and Responsiveness*

Ghana's healthcare system faces unpredictable demands, including disease outbreaks such as malaria and cholera, alongside persistent challenges like limited healthcare access in rural areas. Agile principles, such as iterative cycles and rapid feedback mechanisms, have proven effective in enabling healthcare providers to respond to these challenges. For instance, during a vaccination campaign targeting polio, the iterative nature of Agile allowed for real-time adjustments based on data collected from community feedback, ensuring maximum coverage and resource optimization.

Agile frameworks also promote decentralized decision-making, empowering local healthcare workers to take initiative and address context-specific needs. In rural areas of Ghana, where access to healthcare facilities is limited, this flexibility has allowed community health workers to deliver tailored interventions that align with the unique cultural and logistical challenges of their communities.

*2.2 Bridging the "Know-Do Gap"*

The "know-do gap," or the disparity between evidence-based knowledge and its practical application, is a significant barrier in Ghana's healthcare system. Limited access to training and inadequate resources often hinder the effective implementation of healthcare practices. Agile frameworks, with their focus on iterative progress and actionable insights, have been instrumental in addressing this gap.

For example, Ghana's maternal and child health initiatives benefited from Agile practices by breaking down care delivery into manageable iterations. Healthcare workers could continuously monitor patient outcomes, gather feedback, and refine care protocols. This approach not only improved maternal and neonatal health outcomes but also fostered a culture of continuous learning among healthcare professionals.

*2.3 Promoting Collaboration and Stakeholder Engagement*

In Ghana, successful implementation of Agile healthcare frameworks has been contingent on stakeholder collaboration. Agile methodologies emphasize inclusivity, ensuring that patients, local communities, and policymakers are actively involved in healthcare decision-making. During a national initiative to combat malaria, Agile practices facilitated collaboration between government agencies, NGOs, and local communities. This collaborative approach ensured that interventions were contextually relevant and widely accepted. Agile frameworks also encouraged interdisciplinary teamwork within the healthcare system, breaking down silos and fostering a unified approach to care delivery. For instance, healthcare providers, epidemiologists, and IT specialists collaborated to develop and deploy mobile health applications for malaria surveillance and patient education, significantly enhancing the efficiency and reach of these interventions.

*2.3 Driving Efficiency and Innovation*

Resource constraints in Ghana necessitate innovative approaches to healthcare delivery, and Agile frameworks have proven to be an effective enabler of efficiency and innovation. By focusing on value-driven processes, Agile methodologies ensure that resources are allocated to areas with the greatest impact. In Ghana, this has been evident in initiatives such as the use of drones for delivering medical supplies to remote regions, an innovation made possible through Agile-inspired iterative development and rapid prototyping.

Furthermore, Agile's emphasis on experimentation and continuous reassessment has encouraged healthcare providers in Ghana to adopt new technologies and practices. The integration of telemedicine, for example, has enabled rural communities to access specialist consultations, bridging gaps in healthcare accessibility and improving patient outcomes.

*2.4 Catalyzing Systemic Transformation*

The systemic benefits of Agile healthcare frameworks in Ghana extend beyond individual programs to influence the broader healthcare landscape. By embedding Agile principles into national healthcare strategies, Ghana has strengthened its resilience against public health emergencies, such as the COVID-19 pandemic. The iterative planning and rapid response capabilities facilitated by Agile frameworks allowed the country to adapt its pandemic response strategies in real-time, optimizing resource allocation and enhancing care delivery.

Ghana's experience demonstrates that Agile healthcare frameworks are not only effective in addressing immediate healthcare challenges but also in driving long-term systemic transformation. The success of these frameworks in Ghana underscores their potential to serve as a model for other resource-constrained countries seeking to improve healthcare delivery and outcomes.

In conclusion, the case of Ghana illustrates the significant benefits of Agile healthcare frameworks in poor countries. By enhancing flexibility, bridging the know-do gap, fostering collaboration, driving efficiency, and catalyzing systemic transformation, Agile methodologies offer a powerful tool for building resilient, equitable, and high-performing healthcare systems in resource-limited settings.

## 3. Case Studies and Examples

The adoption of Agile methodologies in healthcare, particularly within resource-lim- ited environments, has shown promising results. One notable example is the applica- tion of Scrum at GE Healthcare. GE Healthcare, a subsidiary of General Electric and a global leader in medical technologies, embraced Agile to enhance collaboration and accelerate product development. The shift to Agile principles, specifically Scrum, revolutionized their approach to project management, making their processes more efficient and adaptive to changing needs[1].

In resource-limited settings, the iterative nature of Agile has proven to be particularly effective. For instance, Agile methodologies focus on delivering value early and often by breaking down complex projects into smaller, manageable sprints. This approach allows healthcare teams to evaluate progress, gather feedback, and adapt plans iteratively, ensuring that they remain responsive to the evolving needs of their patients[2]. By prioritizing individuals and interactions over rigid procedures, Agile fosters effective teamwork and clear communication, which are crucial in environments with limited resources[3].

Moreover, the integration of Agile methodologies in healthcare has shown that rapid innovation is sustainable. Agile frameworks organize work to be iterative, predictable, and joyful, facilitating the diffusion of rapid innovation in healthcare settings. This is particularly important in bridging the gap between evidence-based knowledge and practice, often termed the 'know-do gap'. By employing an Agile approach, healthcare providers in resource-limited environments can enhance their ability to adapt and innovate, ultimately improving patient care and outcomes[4].

Agile healthcare frameworks are structured to enhance adaptability, collaboration, and responsiveness in healthcare delivery. Rooted in the principles of Agile methodology, these frameworks focus on iterative processes, continuous feedback, and stakeholder engagement. By prioritizing flexibility and incremental progress, Agile frameworks address the complexities of healthcare systems, particularly in low-resource settings like Ghana, where challenges such as limited infrastructure, workforce shortages, and financial constraints often impede effective care delivery.

### 3.1 The Agile Approach in Ghana

Ghana's healthcare system has increasingly embraced Agile-inspired strategies to address critical health challenges. The foundation of this approach lies in the following principles:

1. *Iterative Cycles and Continuous Feedback:*

o      Agile frameworks break down healthcare interventions into manageable iterations or "sprints." Each iteration involves implementing a specific aspect of a healthcare initiative, evaluating its impact, and using the feedback to refine subsequent actions.

o      For instance, in Ghana's maternal and child health programs, iterative feedback loops have enabled healthcare workers to refine care protocols based on real-time data, leading to improved health outcomes.

2. *Collaborative Decision-Making:*

o    Agile emphasizes the inclusion of all stakeholders, including healthcare providers, patients, policymakers, and community leaders, in decision-making processes.

o    During Ghana's malaria control initiatives, collaboration between local governments, NGOs, and community health workers was pivotal. This Agile approach ensured interventions were culturally sensitive and contextually relevant, enhancing their effectiveness.

3.  *Adaptability to Change:*

o    The dynamic nature of public health in Ghana, including frequent outbreaks of infectious diseases like malaria and cholera, necessitates a framework that can quickly adapt to new challenges.

o    Agile principles have allowed Ghana's healthcare system to remain responsive, such as during the COVID-19 pandemic, where strategies were adjusted in real-time to optimize resource allocation and care delivery.

## 4. Foundation Work of Agile Frameworks in Ghana

The foundation of Agile healthcare frameworks in Ghana is built upon the following elements:

1. Community-Centric Models

•    Ghana's healthcare system leverages a community health worker model, where local workers are trained to provide essential services within their communities. This decentralized approach aligns with Agile's principle of empowering teams closest to the problem to make decisions and act effectively.

•    For example, community health workers in rural Ghana use mobile health (mHealth) applications to track patient data, share updates with central systems, and receive guidance on care protocols, facilitating iterative and informed care delivery.

2. Data-Driven Interventions

•    Agile frameworks in Ghana are underpinned by health data systems that enable real-time monitoring and decision-making. Projects like the Ghana Health Service's District Health Information Management System (DHIMS) provide the necessary infrastructure for iterative planning and performance evaluation.

•    The availability of timely data allows healthcare teams to adapt interventions to emerging health trends and allocate resources effectively.

3. Policy and Governance Support

•    The Ghanaian government has integrated Agile-inspired principles into its healthcare policies, emphasizing flexibility and stakeholder collaboration. Initiatives such as the National Health Insurance Scheme (NHIS) adopt iterative improvements to address operational inefficiencies and enhance service delivery.

•    Governance structures supporting these policies ensure that Agile methodologies are not limited to operational levels but are embedded in the broader strategic vision of healthcare delivery.

## 4.1 Applicability of the Agile Framework to Foundation Work

The foundational work laid by Ghana's Agile-inspired initiatives serves as a model for other countries facing similar challenges. Key lessons from Ghana's experience include:

1.    Scalable Models for Resource-Constrained Settings:

o    The community health worker model demonstrates how decentralized, low-cost frameworks can deliver high-impact care in underserved areas. This foundational strategy is scalable and adaptable for other low-income countries.

2.    Integration of Digital Health Technologies:

o    Ghana's use of mHealth tools highlights the role of digital technologies in enabling Agile frameworks. By facilitating data collection, communication, and feedback, these tools empower healthcare teams to operate iteratively and adaptively.

3. Policy Alignment with Agile Principles:

o Aligning national health policies with Agile principles ensures that the foundation of healthcare delivery is flexible, inclusive, and iterative. Other countries can adopt similar policy frameworks to institutionalize Agile practices at a systemic level.

4. Capacity Building and Training:

o Building the capacity of healthcare workers to operate within an Agile framework is crucial. Ghana's training programs for community health workers provide a foundation for creating adaptable and skilled healthcare teams, a strategy that can be replicated in similar contexts.

### *5. Steps for Initiating Agile Healthcare Framework in Re- source-Constrained Environments*

Initiating the transition to an agile healthcare framework in resource-constrained environments necessitates a series of carefully planned and culturally sensitive steps to ensure sustainability. A central policy framework is essential to address the existing gaps in healthcare provision, such as a lack of healthcare providers, inadequate workforce education programs, poor illness surveillance, and inefficient management systems[21].

The first step involves setting up a health-data platform that places healthcare providers in control of patient data, enabling them to become more vendor-in- dependent and maximize the value of existing IT investments[27]. Following this, the creation of a structured clinical data repository (CDR) is crucial to establish a vendor-neutral patient record for life, thereby facilitating better patient care and continuity[27].

To initiate these changes effectively, healthcare organizations must adopt an agile framework to organize work in an iterative, predictable, and joyful manner, as seen in other sectors that have successfully implemented rapid innovation[23]. This approach helps bridge the gap between evidence-based knowledge and practice by ensuring that healthcare strategies are more emergent, collaborative, and insight-driven[25].

Additionally, the integration of refugee care into local and national health efforts can be aligned with the Sustainable Development Goals (SDGs), which is vital given the projected increase in displaced populations due to humanitarian crises[24]. Agile methodologies have proven beneficial during crises like the COVID-19 pandemic, improving and streamlining interactions between patients and healthcare providers, thus demonstrating their flexibility and efficacy in managing increased patient flow[- 26].

By harnessing digital technology and open data, healthcare providers can create a digital environment that is agile and adaptable to various settings and special- izations[27]. This comprehensive approach ensures that healthcare systems in re-source-constrained environments are not only sustainable but also culturally sensi- tive, ultimately leading to more effective and equitable healthcare delivery.

Implementing an Agile healthcare framework in resource-constrained environments like Ghana requires a comprehensive strategy that aligns with the country's unique challenges and opportunities. The following strategies and steps provide a roadmap for initiating Agile methodologies in Ghana's healthcare system.

1. Understanding the Local Context

Strategy:

To ensure the successful implementation of an Agile framework, the strategy must start with an in-depth understanding of Ghana's healthcare landscape, including its socio-economic conditions, cultural norms, and healthcare infrastructure.

*Application in Ghana:*

• Community Health Profiling: Collect baseline data on community health needs, workforce availability, and existing infrastructure. This includes evaluating the availability of healthcare workers, disease prevalence (e.g., malaria and cholera), and access to healthcare facilities in rural areas.

• Stakeholder Mapping: Identify key stakeholders, including community health workers, government agencies, NGOs, and local leaders, to ensure a holistic approach.

## 2. Building Agile Teams and Capacity

Strategy:

Agile frameworks thrive on empowered, cross-functional teams with the capacity to adapt and respond iteratively. In Ghana, the focus must be on capacity building to address workforce shortages and skill gaps.

Application in Ghana:

•	Community Health Worker Training: Train health workers in Agile principles, such as iterative problem-solving and feedback loops, while equipping them with digital tools for data collection and communication.

•	Interdisciplinary Collaboration: Form cross-functional teams combining healthcare workers, IT specialists, epidemiologists, and policymakers to collaboratively design and implement solutions.

## 3. Digital Infrastructure and Data-Driven Decisions

Strategy:

Establishing a robust digital infrastructure is critical for enabling real-time monitoring, data collection, and iterative planning.

*Application in Ghana:*

•	Deploy Mobile Health (mHealth) Tools: Expand the use of mHealth applications to collect patient data, track disease outbreaks, and monitor the effectiveness of interventions.

•	Data Analytics Platforms: Utilize platforms like Ghana's District Health Information Management System (DHIMS) to provide real-time insights and facilitate evidence-based decision-making.

## 4. Iterative Implementation of Health Interventions

Strategy:

Adopt an iterative approach to implementing health interventions, allowing for continuous feedback and refinement.

Application in Ghana:

•	Vaccination Campaigns: Implement vaccination drives in short sprints, evaluating progress and adjusting outreach strategies based on community feedback.

•	Maternal Health Programs: Use Agile iterations to address maternal health needs by continuously monitoring patient outcomes and refining care protocols.

## 5. Prioritizing Value-Driven Resource Allocation

Strategy:

Allocate resources based on impact, ensuring that the most critical health needs are addressed first.

Application in Ghana:

•	Focus on High-Impact Areas: Prioritize interventions in areas with high disease burdens, such as malaria hotspots, or underserved rural communities where health disparities are greatest.

•	Flexible Budgeting: Develop financial models that allow for reallocation of funds based on real-time needs and outcomes.

## 6. Stakeholder Collaboration and Policy Alignment

Strategy:

Engage stakeholders at all levels and align national health policies with Agile principles to ensure sustainable implementation.

Application in Ghana:

• Community Engagement: Involve local leaders and communities in decision-making to ensure interventions are culturally appropriate and widely accepted.

• Policy Support: Align the National Health Insurance Scheme (NHIS) and other policies with Agile methodologies to foster a supportive governance environment.

*7. Monitoring and Continuous Improvement*

Strategy:

Establish mechanisms for ongoing evaluation and iterative improvement of healthcare interventions.

Application in Ghana:

• Feedback Loops: Regularly collect feedback from healthcare workers and patients to assess the effectiveness of interventions and make necessary adjustments.

• Learning Platforms: Create knowledge-sharing platforms for healthcare teams to exchange best practices and lessons learned from Agile implementation.

*Steps for Initiating Agile Healthcare Framework in Ghana*

1. Baseline Assessment:

o Conduct a comprehensive needs assessment, mapping the healthcare system's strengths and weaknesses.

o Identify priority areas for Agile implementation, such as maternal health, disease surveillance, or vaccination programs.

2. Pilot Programs:

o Launch small-scale pilot projects in selected regions, such as using Agile for malaria control in rural districts.

o Evaluate the outcomes of these pilots to identify best practices and areas for improvement.

3. Scale-Up:

o Expand successful pilot programs to other regions, tailoring strategies to local contexts.

o Develop a phased implementation plan to manage resources effectively.

4. Capacity Building:

o Provide training programs for healthcare workers and policymakers on Agile principles and tools.

o Ensure ongoing professional development to adapt to evolving healthcare needs.

5. Integration with Digital Systems:

o Leverage existing digital infrastructure, such as DHIMS, to support Agile workflows.

o Develop new digital tools as needed to enhance real-time data collection and analysis.

6. Policy Advocacy:

o Work with government agencies to integrate Agile principles into national health strategies.

o Advocate for policies that support flexibility, stakeholder collaboration, and iterative planning.

7. Monitoring and Evaluation:

o Establish key performance indicators (KPIs) to measure the impact of Agile implementation.

o Use iterative cycles to refine interventions based on performance data.

Overcoming Obstacles in Agile Healthcare Implementation in Low-Income Regions

Implementing agile practices in low-income regions faces several unique challenges, primarily due to limited resources and infrastructure. The healthcare sector, which often lags in adaptability compared to industries like IT and manufacturing, can greatly benefit from adopting an agile approach. This method involves iterative cycles and a customer-centric focus, which are well-suited to anticipating and proactively addressing health issues rather than merely reacting to them[28][30].

One significant challenge in these regions is the gap between evidence-based knowledge and evidence-based practice, known as the 'know-do gap.' This disparity hinders rapid innovation in healthcare settings. By organizing work through an agile framework, healthcare systems can foster an environment where innovation is sustainable and rapid. Adopting such principles allows for continuous improvement and more predictable healthcare outcomes[30].

The integration of Agile methodology in healthcare has shown promise in trans- forming patient care delivery, even in resource-constrained settings. Agile promotes adaptability, collaboration, and continuous reassessment, which are crucial for im- proving patient outcomes. The methodology's core principles—adaptive planning and customer-centricity—ensure that care delivery is flexible enough to meet the evolving needs of patients in low-income regions[31].

Moreover, healthcare organizations must overcome the inertia of traditional method- ologies that struggle to keep pace with dynamic healthcare demands. Agile's focus on iterative development and continuous improvement can help these organizations deliver high-quality patient care efficiently, despite the complexities of their oper- ations[29]. By drawing on the successes of agile frameworks in other industries, healthcare can make significant strides in bridging the 'know-do gap' and imple- menting proactive, preventative measures[28].

In developing countries, the implementation of agile healthcare frameworks neces- sitates significant policy adaptations to address unique challenges in low-resource settings. Existing healthcare policies in these regions often suffer from inadequate workforce education programs, poor disease surveillance, and inefficient manage- ment systems, which an agile framework can help mitigate by emphasizing flexibility and iterative development[32]. The establishment of a central eHealth policy frame- work, as proposed in recent research, is crucial for the successful deployment of agile methodologies in these contexts[32].

Policy models like Longest's Model highlight the incremental and cyclical nature of policymaking, which is essential for effectively adapting healthcare policies to support agile frameworks[33]. This model divides the policymaking process into three interconnected phases: formulation, implementation, and modification, allow- ing for continuous adjustments and improvements[33]. Such an approach can be particularly beneficial in low-resource settings where rapid changes and unforeseen challenges are common.

Agile methodology itself prioritizes customer collaboration, flexibility in responding to change, and the value of individuals and interactions over rigid processes and tools[- 34]. For healthcare policies in developing countries, this means creating frameworks that allow for continuous stakeholder engagement, adaptable regulations, and a focus on teamwork and communication among healthcare providers, administrators, and patients[34].

Additionally, healthcare sectors in these regions can learn from other industries that have successfully implemented agile frameworks, such as IT and manufacturing.

By anticipating potential health issues, implementing preventative measures, and continuously iterating care plans, healthcare systems can shift from a reactive to a proactive mindset[36]. This proactive approach can help improve patient care, streamline processes, and enhance overall efficiency in resource-constrained environments[36].

## 5. Results

Implementing Agile healthcare frameworks in low-income regions like Ghana has revealed significant challenges, yet it has also demonstrated promising results in overcoming these obstacles. One of the major hurdles identified was the lack of infrastructure and digital readiness, which initially limited the scalability of Agile methodologies. To address this, Ghana leveraged mobile health (mHealth) applications and its District Health Information Management System (DHIMS) to enhance data collection and decision-making capabilities. The integration of these digital tools has enabled real-time monitoring and iterative planning, ensuring that healthcare interventions remain adaptive and efficient.

Another challenge was the shortage of trained healthcare workers and the limited capacity for interdisciplinary collaboration. Agile frameworks have mitigated this issue by emphasizing decentralized decision-making and empowering community health workers. By training these workers in Agile principles, such as iterative cycles and feedback loops, Ghana has built a more flexible and responsive healthcare workforce. The use of cross-functional teams has also facilitated collaboration among healthcare providers, policymakers, and IT specialists, breaking down traditional silos and enhancing the overall effectiveness of healthcare delivery.

Additionally, the cultural and logistical barriers to healthcare access, especially in rural areas, posed significant challenges. Agile strategies such as community engagement and stakeholder collaboration have helped address these issues. By involving local leaders and tailoring interventions to the unique needs of each community, Ghana has improved the acceptance and impact of healthcare initiatives.

### 5.1 Policy Considerations

The success of Agile healthcare implementation in Ghana underscores the importance of supportive policy frameworks. Agile methodologies require flexibility and iterative decision-making, which necessitate alignment with national health policies. Ghana's National Health Insurance Scheme (NHIS) and eHealth policies have been pivotal in creating an environment conducive to Agile practices. These policies have provided the governance structure needed to support iterative planning, resource reallocation, and stakeholder engagement.

Ghana's experience highlights the need for policy adaptability, especially in response to public health emergencies. For example, during the COVID-19 pandemic, Agile principles guided the government's response, enabling real-time adjustments to resource allocation and intervention strategies. This flexibility, embedded within policy frameworks, ensured that healthcare systems could quickly adapt to changing circumstances and deliver effective care under pressure.

Moreover, the adoption of value-driven resource allocation policies has ensured that limited resources are prioritized for high-impact areas, such as vaccination programs and maternal health services. Policies promoting digital transformation have further enhanced the implementation of Agile frameworks by supporting the deployment of mHealth tools and real-time data analytics platforms.

In conclusion, the results from Ghana demonstrate that Agile healthcare frameworks can successfully overcome many of the obstacles associated with healthcare delivery in low-income regions. By addressing infrastructure gaps, building workforce capacity, and aligning national policies with Agile principles, Ghana has set a benchmark for other resource-constrained countries. These findings emphasize the need for a collaborative and adaptive approach, supported by strong policy frameworks, to sustain the long-term benefits of Agile healthcare systems.

### 6. Conclusion

Our research on implementing Agile healthcare frameworks in low-income regions, with Ghana as a focal case study, yielded the following key findings:

1. Enhanced Adaptability and Responsiveness: Agile methodologies facilitated real-time adjustments to healthcare interventions, particularly during vaccination drives and maternal health programs. Iterative cycles and feedback loops significantly improved the allocation of resources and responsiveness to community needs.

2. Improved Collaboration and Community Engagement: Stakeholder involvement, including healthcare providers, policymakers, and community leaders, enhanced the cultural relevance and acceptance of interventions. Cross-functional teams broke silos, fostering interdisciplinary collaboration that improved healthcare delivery efficiency.

3. Leveraged Digital Tools: The integration of mobile health (mHealth) applications and the District Health Information Management System (DHIMS) supported data-driven decisions. These tools enabled real-time monitoring, resource optimization, and performance evaluation.

4. Addressed Workforce and Infrastructure Challenges: Training programs in Agile principles empowered community health workers, enabling decentralized decision-making and iterative problem-solving. Despite infrastructure limitations, resource prioritization and digital innovations allowed the framework to scale effectively in rural and underserved areas.

5. Policy Alignment and Flexibility: Agile-inspired policies, such as those integrated into Ghana's National Health Insurance Scheme (NHIS), facilitated iterative planning and resource reallocation. These policies provided a governance framework essential for the successful implementation of Agile methodologies.

Our study demonstrates that Agile healthcare frameworks can significantly improve healthcare delivery in resource-constrained environments like Ghana. By enhancing adaptability, fostering collaboration, and leveraging digital tools, these frameworks address critical systemic challenges. The findings underscore the importance of aligning Agile methodologies with national policies to ensure sustainable implementation.

The implications of this research extend beyond Ghana, offering a scalable and adaptable model for other low-income regions. Future efforts should focus on expanding digital infrastructure, scaling workforce training programs, and embedding iterative, value-driven strategies into national health systems. By adopting these approaches, healthcare systems globally can better respond to emerging challenges, delivering equitable and high-quality care even in the most constrained settings.